\documentclass[prb,showpacs,floatfix,twocolumn]{revtex4}
\usepackage{graphicx}

\begin{document}

\title{Structural distortion below the N\'eel temperature in spinel GeCo$_2$O$_4$} 

\author{Phillip T. Barton}
\author{Moureen C. Kemei}\email{kemei@mrl.ucsb.edu}
\author{Michael W. Gaultois}
\author{Stephanie L. Moffitt}
\author{Lucy E. Darago}
\author{Ram Seshadri} \email{seshadri@mrl.ucsb.edu}
\affiliation{Materials Department and Materials Research Laboratory, University of California, Santa Barbara, CA, 93106, USA}

\author{Matthew R. Suchomel}
\affiliation{X-Ray Science Division, Argonne National Laboratory, Argonne, IL 60439, USA}

\author{Brent C. Melot}
\affiliation{Department of Chemistry, University of Southern California, Los Angeles, CA, USA}

\date{\today}

\begin{abstract}

A structural phase transition from cubic $Fd\bar{3}m$ to tetragonal $I$4$_1$/$amd$ symmetry with $c/a >$\,1 is observed at $T_{\rm{S}}$~=~16\,K in spinel GeCo$_2$O$_4$ below the N\'eel temperature $T_N$\,=\,21\,K. Structural and magnetic ordering appear to be decoupled with the structural distortion occurring at 16\,K while magnetic order occurs at 21\,K as determined by magnetic susceptibility and heat capacity measurements. An elongation of CoO$_6$ octahedra is observed in the tetragonal phase of GeCo$_2$O$_4$. We present the complete crystallographic description of GeCo$_2$O$_4$ in the tetragonal $I$4$_1$/$amd$ space group and discuss the possible origin of this distortion in the context of known structural transitions in magnetic spinels. GeCo$_2$O$_4$ exhibits magnetodielectric coupling below $T_{\rm{N}}$. The related spinels GeFe$_2$O$_4$ and GeNi$_2$O$_4$ have also been examined for comparison. Structural transitions were not detected in either compound down to $T\,\approx$\,8\,K. Magnetometry experiments reveal in GeFe$_2$O$_4$ a second antiferromagnetic transition, with $T_{\rm{N1}}$~=~7.9\,K and $T_{\rm{N2}}$~=~6.2\,K, that was previously unknown, and that bear a similarity to the magnetism of GeNi$_2$O$_4$.


\end{abstract}

\maketitle 

\section{Introduction} 

The spinel crystal structure is of wide interest in condensed matter physics for diverse phenomena including heavy fermions,\cite{Kondo_PRL97} multiferroic behavior,\cite{Yamasaki_PRL06} and exotic states arising from geometric frustration.\cite{Bordacs_PRL09,LaForge_PRL13,Kemei_JPCM13} The rich physics of complex transition metal oxides derives from the intricate interplay of charge, orbital, spin, and lattice degrees of freedom. In this report, we examine the magnetic and structural properties of the spinel GeCo$_2$O$_4$ that are largely influenced by competing orbital and spin degrees of freedom. We also study the structure and magnetism of the related systems Ge$M_2$O$_4$ ($M$ = Fe and Ni).


At room temperature, Ge$M_2$O$_4$ ($M$ = Fe, Co, and Ni) are cubic spinel oxides in the space group $Fd\overline{3}m$. Ge$^{4+}$ cations are tetrahedrally coordinated by O$^{2-}$ while $M^{2+}$ cations occupy octahedral sites. The structures and magnetic behavior of all three Ge$M_2$O$_4$ spinels were originally reported by Blasse and Fast in 1963.\cite{Blasse_PRR63,Gorter_JAP63} GeFe$_2$O$_4$ is orbitally degenerate due to partially filled $t_{2g}^4$ states of octahedral high spin Fe$^{2+}$. GeCo$_2$O$_4$ has been the subject of many investigations because it has the unique electronic ground state of octahedral Co(II), which is high-spin 3$d^7$, with $S$ = 3/2 $L$ = 3, though it is better described as a Kramer's doublet with $J_{\rm{eff}}$ = 1/2. The orbitally degenerate $t_{2g}^5$ states of high spin octahedral Co$^{2+}$ give rise to spin-orbit coupling that results in a large single-ion anisotropy for a 3$d$ transition metal. In contrast to GeCo$_2$O$_4$ and GeFe$_2$O$_4$, GeNi$_2$O$_4$ has a non-degenerate electronic ground state with fully occupied $t_{2g}^6$ levels and half occupied $e_g^2$ states of octahedral Ni$^{2+}$.

GeCo$_2$O$_4$, GeFe$_2$O$_4$, and GeNi$_2$O$_4$ order antiferromagnetic at temperatures below 30\,K. GeCo$_2$O$_4$ has a N\'eel temperature near 21\,K while GeNi$_2$O$_4$ shows two magnetic ordering anomalies at $\approx$\,12\,K and 11\,K.\cite{Lashley_PRB08,Diaz_PRB06,Crawford_PRB03} Our magnetic susceptibility studies of the spinel GeFe$_2$O$_4$ show that it also exhibits two antiferromagnetic transitions at 7.9\,K and 6.2\,K. The magnetic structure of the Ni and Co compounds consists of ferromagnetic (111) planes that are antiferromagnetically coupled with a ($\frac{1}{2}$ $\frac{1}{2}$ $\frac{1}{2}$) magnetic propagation vector.\cite{Diaz_PRB06} In between the Kagome planes are triangular planes of spins whose orientation is not well known. Neutron diffraction measurements by Diaz $et\,al.$ show that the triangular plane moments of GeNi$_2$O$_4$ are aligned parallel to the (111) direction while in GeCo$_2$O$_4$ the triangular plane moments are perpendicular to the (111) direction.\cite{Diaz_PRB06} Diaz $et\,al.$ have also shown that GeCo$_2$O$_4$ and GeNi$_2$O$_4$ systems undergo two subtle field-induced transitions above 4\,T.\cite{Diaz_PRB06} 

Here, we study the low temperature tetragonal structural distortion of the spinel GeCo$_2$O$_4$. We find that the structural distortion is decoupled from antiferromagnetic ordering, occurring at $T_D$\,=\,16\,K rather than at the N\'eel temperature of 21\,K. We resolve the low-temperature nuclear structure of GeCo$_2$O$_4$ by Rietveld refinement of high resolution synchrotron x-ray diffraction data using a tetragonal $I$4$_1$/$amd$ model with $c/a >$ 1. The evolution of structure shows an elongation of CoO$_6$ octahedra in the tetragonal phase of GeCo$_2$O$_4$. We discuss the mechanisms behind the structural distortion of GeCo$_2$O$_4$ in the context of known structural distortions in magnetic spinels. Synchrotron diffraction studies of GeFe$_2$O$_4$ and GeNi$_2$O$_4$ down to $\approx$\,8\,K show the absence of structural distortions in these systems above this temperature. We also report magnetodielectric coupling in GeCo$_2$O$_4$ beneath $T_{\rm{N}}$~=~23\,K, while GeNi$_2$O$_4$ shows no evidence for such behavior. Magnetic susceptibility studies of the related spinel, GeFe$_2$O$_4$, reveals two antiferromagnetic ordering temperatures of 6.2\,K and 7.9\,K.

\section{Methods}

Polycrystalline Ge$M_2$O$_4$ ($M$ = Fe, Co, Ni) were prepared by solid-state reaction of powder reagents. Stoichiometric amounts of GeO$_2$ and either Fe/Fe$_2$O$_3$, Co$_3$O$_4$, or NiO were ground with an agate mortar and pestle and pressed into pellets at a pressure of 100\,MPa. The pellet of the Fe compound was sealed inside an evacuated quartz ampoule to maintain the oxygen stoichiometry necessary for Fe(II). The Co and Ni compound pellets were placed inside Al$_2$O$_3$ crucibles on top of a bed of powder with the same composition in order to avoid contamination from the crucible. The sealed tube of the Fe compound was heated to 800$^{\circ}$C, while the Co compound was annealed at 1000$^{\circ}$C. The reactions occurred in a box furnace for two days with one intermediate grinding and repressing of the powder. The preparation of GeNi$_2$O$_4$ involved heating the loose powder slowly to 900$^{\circ}$C and annealing for 12 hours, followed by grinding, pelletization, and annealing at 1100$^{\circ}$C for 24 hours and at 1200$^{\circ}$C for another 24 hours. Powder synchrotron x-ray diffraction was conducted at both the 11-BM beamline ($\lambda$ $\approx$ 0.41317\,\AA) of the Advanced Photon Source, Argonne National Laboratory and the ID31 beamline ($\lambda$ $\approx$ 0.399845\,\AA) of the European Synchrotron Radiation Facility. Powder coated Kapton capillaries were employed to reduce synchrotron x-ray beam heating and improve temperature equilibrium with the closed Helium cryostat exchange gas. During the study of GeCo$_2$O$_4$, the temperature was varied at 0.05\,K/min in the temperature range 6.6\,K\,$<$\,$T$\,$<$\,24\,K, and an x-ray scan was measured every 5 minutes. The temperature difference during the course of a given scan in this temperature regime was 0.25\,K. A faster temperature ramp rate of 1\,K/min was applied in the temperature range 28\,K\,$<$\,$T$\,$<$\,60\,K and a x-ray scan was measured every 2.5\,min. Variable-temperature x-ray measurements of GeNi$_2$O$_4$ were measured at 0.5\,K/min and an x-ray scan was measured every 5 minutes over the temperature range 7.5\,K\,$<$\,$T$\,$<$\,130\,K. GeFe$_2$O$_4$ was studied at 2\,K/min, with an x-ray scan being measured every 3\,minutes in the temperature range 7.7\,K\,$<$\,$T$\,$<$\,130\,K.  Separate low temperature synchrotron x-ray measurements of GeFe$_2$O$_4$, GeCo$_2$O$_4$, and GeNi$_2$O$_4$ down to 5\,K were performed at the European Synchrotron Radiation Facility. Rietveld\cite{Rietveld_JAC1969} analyses were performed using GSAS/EXPGUI.\cite{Toby_JAC2001}  DICVOL, as implemented in FullProf, was used to index the low-temperature unit cell.\cite{Boultif_JAC2004} ISODISTORT was used to explore the possible crystal distortion modes and to transform the unit cell atom positions to lower symmetry.\cite{Campbell_JAC2006} Crystal structures were visualized using VESTA.\cite{Momma_VESTA_2008} Magnetic properties were measured using a Quantum Design MPMS 5XL SQUID magnetometer. Capacitance was measured using a 1\,V excitation in a parallel plate geometry with an Andeen-Hagerling bridge in a Quantum Design PPMS DynaCool cryostat. Prior to measurement, capacitance samples were densified through spark plasma sintering and coated with silver epoxy paste for electrodes. The processing did not affect the material crystal structure or composition, as determined by synchrotron x-ray diffraction.

\begin{figure*}
\centering
\includegraphics[width=6in]{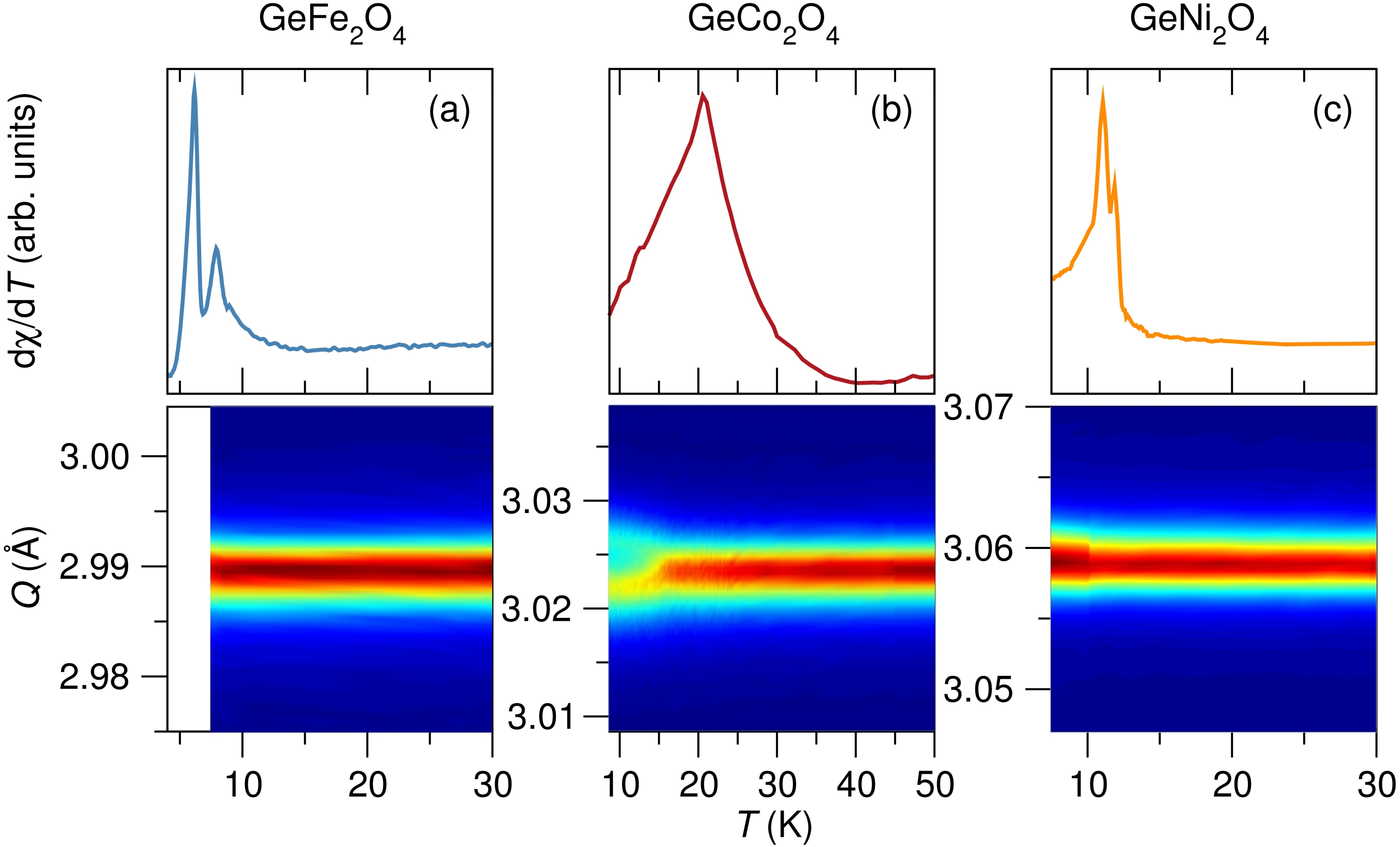}
\caption{(Color online) GeFe$_2$O$_4$, GeCo$_2$O$_4$, and GeNi$_2$O$_4$ are antiferromagnetic and the derivative of magnetic susceptibility as a function of temperature shows sharp anomalies at their respective N\'eel temperatures. (a) GeFe$_2$O$_4$ shows two magnetic ordering temperatures at 6.2\,K and 7.9\,K. The N\'eel temperature of (b) GeCo$_2$O$_4$ is 20.9\,K while (c) GeNi$_2$O$_4$ orders at 11.1\,K and 11.9\,K. Variable-temperature high-resolution synchrotron x-ray powder diffraction of GeFe$_2$O$_4$, GeCo$_2$O$_4$, and GeNi$_2$O$_4$ shows that magnetic ordering in these spinels is decoupled from the structure(bottom panel). No structural transitions are observed in GeFe$_2$O$_4$ and GeNi$_2$O$_4$ where the (400) cubic $\textit{Fd}\overline{3}\textit{m}$ reflection shows no splitting in the temperature range 8\,K\,$\leq$\,$T$\,$\leq$\,30\,K. In contrast, the (400) cubic $\textit{Fd}\overline{3}\textit{m}$ reflection of GeCo$_2$O$_4$ splits into (004) and (220) tetragonal $I$4$_1$/$\textit{amd}$ reflections at 16\,K rather than at the N\'eel temperature. }
\label{fig:vt}
\end{figure*}



\section{Results and Discussion}

The crystal structures of powder GeFe$_2$O$_4$, GeCo$_2$O$_4$, and GeNi$_2$O$_4$ samples were investigated by synchrotron x-ray powder diffraction in the temperature range 6.6\,K\,$\lesssim$\,$T$\,$\leq$\,295\,K. Unit cell parameters at $T$\,=\,295\,K of $a_{\rm{Fe}}$~=~8.41368(8)\,\AA\/, $a_{\rm{Co}}$~=~8.31910(8)\,\AA\/,  and $a_{\rm{Ni}}$~=~8.22422(4)\,\AA\/ were extracted by Rietveld refinement of x-ray data and are in accord with prior investigations.\cite{Welch_MM01,Furuhashi_JINC73,Hirota_JCSJ90} The known room-temperature spinel crystal structure was determined by Rietveld refinement of the diffraction pattern using the space group $Fd\bar{3}m$. A small impurity phase was detected in the GeFe$_2$O$_4$ sample and was determined to be 5.4\,wt$\%$ of Fe$_{1.67}$Ge.\cite{Kanematsu_JPSJ65}  A Co$_{10}$Ge$_3$O$_{16}$ impurity at a level of 1.4\,wt$\%$ was identified in GeCo$_2$O$_4$. The impurities Fe$_{1.67}$Ge and Co$_{10}$Ge$_3$O$_{16}$, whose properties are reported by Barbier\cite{Barbier_ACC1995} and Barton $et\,al.$\cite{Barton_PRB13} respectively, have a minor influence on the results. Bond valence sum calculations are consistent with the 2+ valence state for each of these transition metal ions. 

Magnetic susceptibility measurements of GeFe$_2$O$_4$, GeCo$_2$O$_4$, and GeNi$_2$O$_4$ show that they are antiferromagnetic at low temperature. Figure \ref{fig:vt} shows the d$\chi$/d$T$ curves for these spinels. Peaks in d$\chi$/d$T$ of GeFe$_2$O$_4$ occur at both $T_{\rm{N1}}$~=~7.9\,K and $T_{\rm{N2}}$~=~6.2\,K showing evidence for a second antiferromagnetic transition in GeFe$_2$O$_4$ that has not been reported [Fig. \ref{fig:vt} (a)]. This behavior is similar to that of GeNi$_2$O$_4$, which is known to exhibit two transitions\cite{Crawford_PRB03} that we observe at 11.9\,K and 11.1\,K [Fig. \ref{fig:vt} (c)]. A neutron diffraction study by Matsuda \textit{et al.} attributes the two transitions of GeNi$_2$O$_4$ to separate orderings of the spins in the Kagome and triangular planes.\cite{Matsuda_EPL08} Curie-Weiss fitting of the high temperature susceptibility of GeNi$_2$O$_4$ leads to $\mu_{\rm{eff}}$~=~3.36\,$\mu_{\rm{B}}$ and $\Theta_{\rm{CW}}$~=~$-$11.3\,K, congruent with the literature for GeNi$_2$O$_4$.\cite{Diaz_PB04} A cusp in the d$\chi$/d$T$ of GeCo$_2$O$_4$ at $T_{\rm{N}}$~=~20.9\,K indicates the onset of long-range antiferromagnetic order [Fig. \ref{fig:vt} (b)], consistent with previous reports on GeCo$_2$O$_4$. Though it is not strictly valid to apply Curie-Weiss to GeCo$_2$O$_4$ because of Co(II) crystal field levels,\cite{Lashley_PRB08} we find $\mu_{\rm{eff}}$ = 4.55\,$\mu_{\rm{B}}$ and $\Theta_{\rm{CW}}$ = 55.0\,K, in reasonable agreement with the literature.\cite{Diaz_PB04} We are unable to analyze the magnetic susceptibility of GeFe$_2$O$_4$ by Curie-Weiss analysis because of the ferromagnetic Fe$_{1.67}$Ge impurity with $T_{\rm{C}}$~=~485\,K.\cite{Yasuk_JPSJ61} 

Variable-temperature synchrotron x-ray powder diffraction patterns show no evidence of a structural phase transition in either GeFe$_2$O$_4$ nor in GeNi$_2$O$_4$ down to $T$\,=\,8\,K [bottom panel of Fig. \ref{fig:vt}]. A slight broadening of the cubic (400) $\textit{Fd}\overline{3}\textit{m}$ reflection occurs in GeNi$_2$O$_4$ at the N\'eel temperature but a splitting of the reflection is not observed. Near 8\,K, GeFe$_2$O$_4$ and GeNi$_2$O$_4$ are well modeled by the cubic $\textit{Fd}\overline{3}\textit{m}$ structure and we determine the unit cell parameters $a_{\rm{Fe}}$~=~8.40508(1)\,\AA\/ and $a_{\rm{Ni}}$~=~8.21569(2)\,\AA\/. Separate measurements show that GeFe$_2$O$_4$ and GeNi$_2$O$_4$ retain cubic symmetry even at 5\,K. The unique electronic configuration of octahedral Ni$^{2+}$ $t_{2g}^6$ $e_g^2$ in GeNi$_2$O$_4$ precludes the presence of any Jahn-Teller activity and previous studies of this material also found no evidence of a magnetostructural distortion.\cite{crawford_2003} GeFe$_2$O$_4$ has not been extensively studied and our measurements show no structural distortions from cubic symmetry even at 5\,K, although Fe$^{2+}$ cations are orbitally degenerate with partially filled $t_{2g}^4$ states. In contrast to GeFe$_2$O$_4$ and GeNi$_2$O$_4$, the (400) cubic $\textit{Fd}\overline{3}\textit{m}$ reflection of GeCo$_2$O$_4$ splits at $T_D$\,$\approx$\,16\,K [bottom panel of Fig. \ref{fig:vt}], confirming the onset of its known structural phase transition at low temperatures.\cite{Hoshi_JMMM07} Figure \ref{fig:vt} shows a discrepancy between the onset of antiferromagnetic order in GeCo$_2$O$_4$ at $T_N$\,$\approx$\,21\,K [Fig. \ref{fig:vt} (b)] and the onset of the structural distortion at $T_D$\,$\approx$\,16\,K.

\begin{figure}
\centering \includegraphics[width=3.4in]{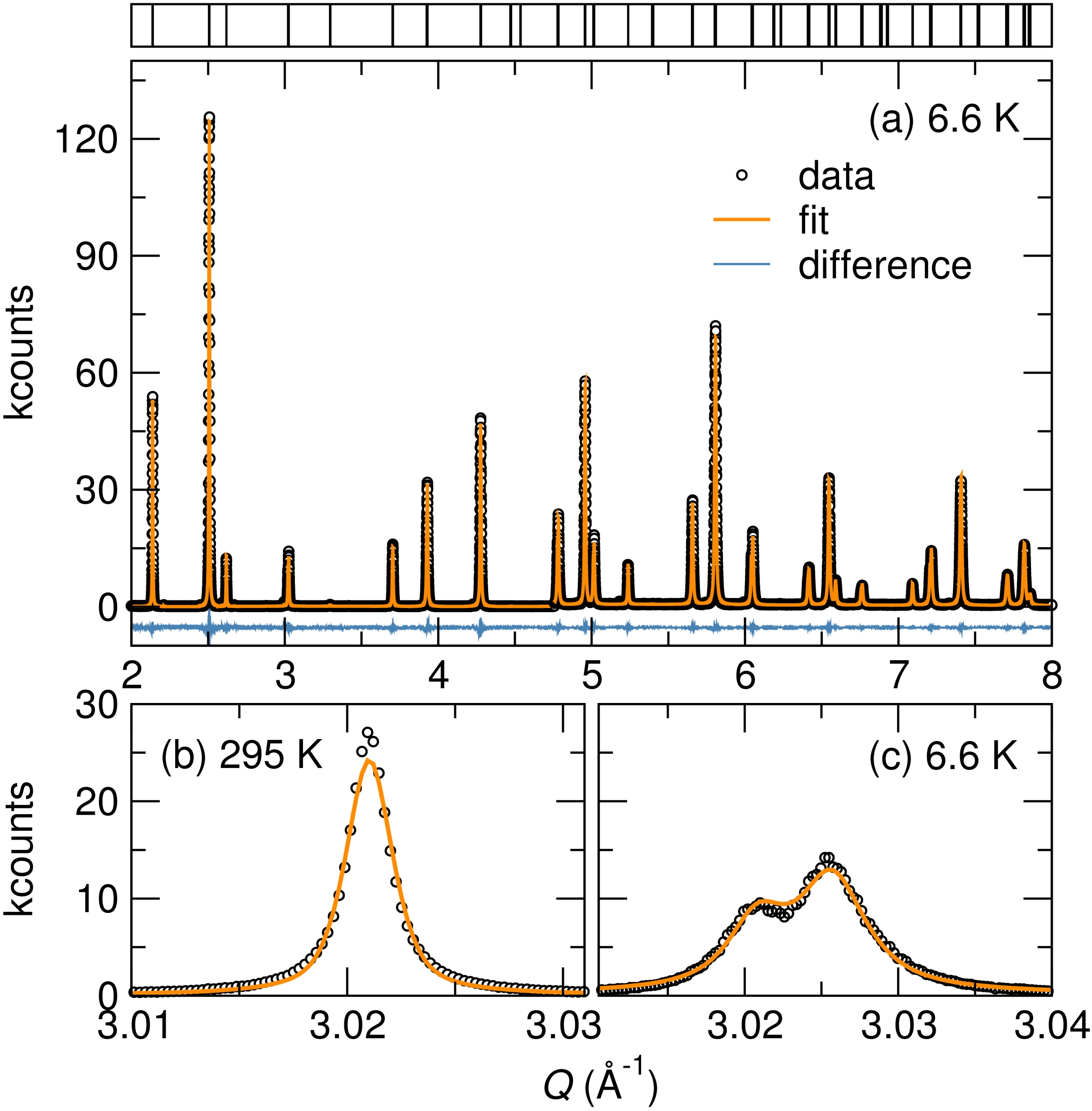}\\
\caption{(Color online) (a) Powder synchrotron x-ray diffraction of GeCo$_2$O$_4$ at $T$ = 6.6\,K modeled by Rietveld refinement to a tetragonal $\textit{I}$4$_1$/$\textit{amd}$ structure and 1.4\,wt$\%$ Co$_{10}$Ge$_3$O$_{16}$ impurity phase. Note that data are multiplied by 5 for $Q$\,$>$\,4.75\,\AA\, for visual clarity. The cubic (400) $\textit{Fd}\overline{3}\textit{m}$ reflection of GeCo$_2$O$_4$ shown in (b) splits to two tetragonal (004) and (220) reflections that are modeled by the $\textit{I}$4$_1$/$\textit{amd}$ structure (c).
\label{fig:Rietveld}}
\end{figure}
\begin{table}
\caption{Structural parameters of GeCo$_2$O$_4$ at $T$ = 6.6\,K. Space group: $I$4$_1$/$amd$, $a$ = 5.87338(1)\,\AA\, and $c$ = 8.31957(2)\,\AA. The refinement figures of merit of $R_{\rm{wp}}$ and $R_{\rm{p}}$ are 4.34\,\% and 8.68\,\% respectively.
\label{tab:I41/amd}}
\centering
\begin{tabular}{lccccccccccccc}
\hline
\hline
Site &$x$ &$y$ &$z$ &$U_{iso}$ (\AA$^2$) \\
\hline
Ge &0 &0.25 &0.375 &0.0020(1)\ \\
Co &0 &0 &0 &0.0027(1)\ \\
O &0 &0.5010(1) &0.2519(1) & 0.0021(1)\ \\
\hline
\hline
\end{tabular}
\end{table}

We quantitatively describe the low-temperature synchrotron x-ray powder diffraction pattern of GeCo$_2$O$_4$ with a tetragonal $\textit{I}$4$_1$/$\textit{amd}$ model which is a subgroup of the $\textit{Fd}\overline{3}\textit{m}$ space group that is commonly used to describe other spinel systems that undergo structural distortions from cubic $Fd\bar{3}m$ symmetry.\cite{Suchomel_PRB12} The initial unit cell parameters for the tetragonal model were determined by diffraction pattern indexing and its atom positions were derived using group-subgroup theory. Figure~\ref{fig:Rietveld} (a) displays the refinement of the $T$ = 6.6\,K experimental data for GeCo$_2$O$_4$ to the $\textit{I}$4$_1$/$\textit{amd}$ model. The cubic (400) reflection [Fig. \ref{fig:Rietveld} (b)] splits in the structurally distorted phase as shown in fig. \ref{fig:Rietveld} (c) and this divergence of the diffraction reflection is well described by the $\textit{I}$4$_1$/$\textit{amd}$  model. The small difference between the data and the structural model [Fig. \ref{fig:Rietveld} (a)] and  refinement figures of merit [Table \ref{tab:I41/amd}] support the validity of the low-temperature tetragonal $\textit{1}$4$_1$/$\textit{amd}$ structural model. The extracted structural parameters for the $\textit{I}$4$_1$/$\textit{amd}$ tetragonal structure of GeCo$_2$O$_4$ at $T$~=~6.6\,K are listed in Table~\ref{tab:I41/amd}. 

\begin{figure}
\centering \includegraphics[width=3.4in]{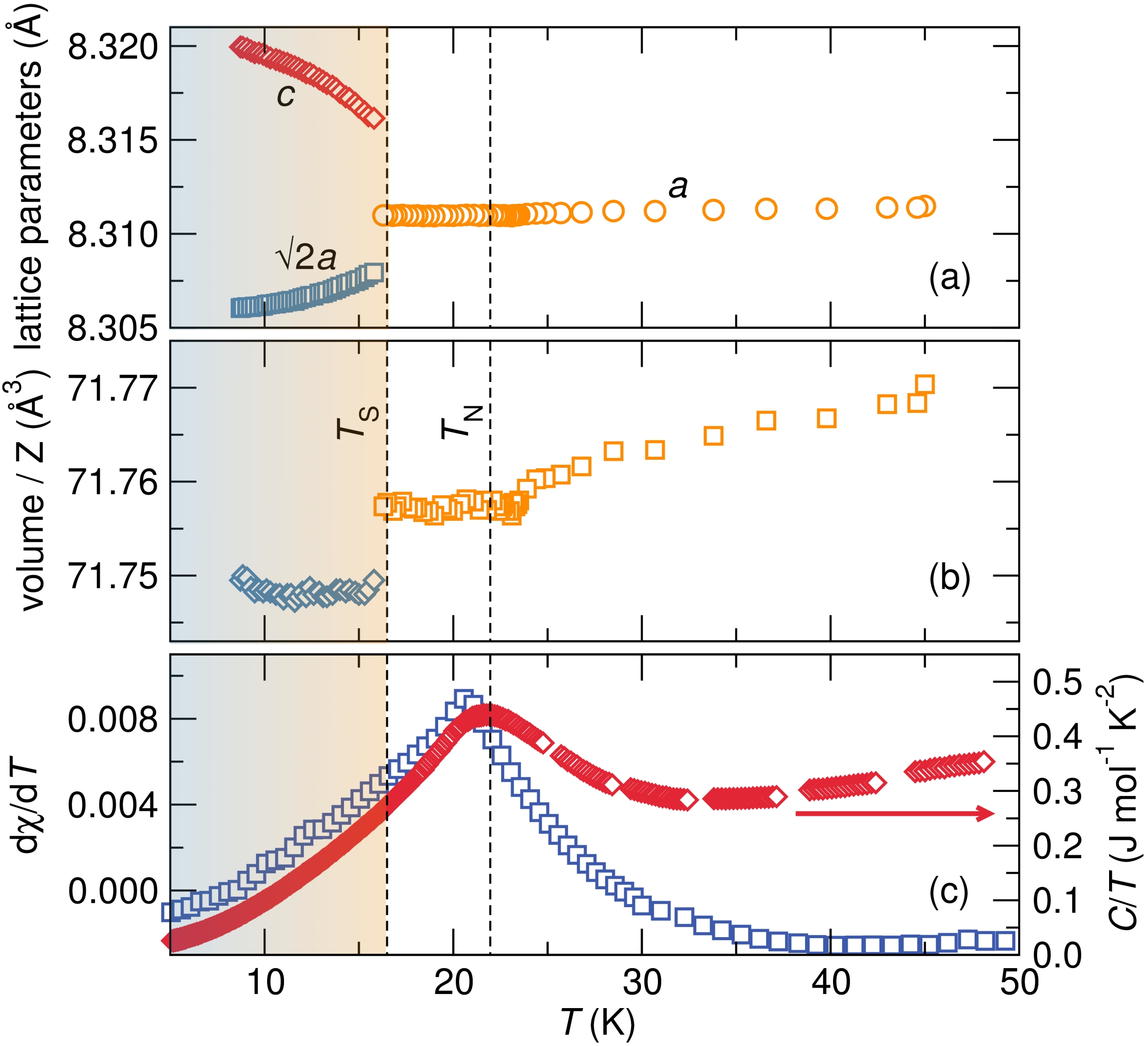}\\
\caption{(Color online) (a) At the structural distortion temperature of GeCo$_2$O$_4$, $T_D$\,=\,16\,K, two tetragonal lattice constants emerge from the cubic lattice constant. An elongation of the tetragonal $c$ axis is observed. (b) The thermal evolution of the cell volume of GeCo$_2$O$_4$ shows two anomalies, one at the antiferromagnetic ordering temperature, $T_N$\,=\,20.9\,K, the other at the structural distortion temperature, $T_D$\,=\,16\,K. (c) d$\chi$/d$T$ and temperature normalized heat capacity measurements show peaks at the antiferromagnetic ordering temperature of GeCo$_2$O$_4$. 
\label{fig:Structureevol}}
\end{figure}

We separately fit the low-temperature tetragonal $\textit{I}$4$_1$/$\textit{amd}$ model and the high temperature cubic  $\textit{Fd}\overline{3}\textit{m}$ structure to the GeCo$_2$O$_4$ diffraction patterns in the temperature region around the transition to determine the structural phase transition temperature of GeCo$_2$O$_4$. Upon examining the stability of the refinements and comparing their figures of merit, the structural transition was determined to occur at $T_{\rm{D}}$~=~16\,K. Two tetragonal lattice constants emerge below 16\,K [Fig. \ref{fig:Structureevol} (a)]. The tetragonal phase is characterized by $c/a\,>$ 1 and the degree of tetragonality, increases with decreasing temperature. For both the high- and low- temperature structures, bond valence sum calculations based on Shannon-Prewitt effective ionic radii,\cite{Shannon_ACB69} indicate the ion valences expected from the stoichiometric chemical formula, namely Ge$^{4+}$, Co$^{2+}$, and O$^{2-}$. The onset of the structural distortion below the N\'eel temperature,  $T_{\rm{N}}$~=~21\,K, is unusual in comparison to our investigations of magnetostructural phase transitions in the $A$Cr$_2$O$_4$ spinels\cite{Kemei_JPCM13,Suchomel_PRB12} which show concurrent magnetic and structural transitions. 

The unit cell volume of GeCo$_2$O$_4$ decreases with temperature, as expected for a material with a positive coefficient of thermal expansion [Fig.~\ref{fig:Structureevol}(b)]. Discontinuities in the cell volume occur at the antiferromagnetic ordering temperature, $T$\,=\,21\,K, due to isotropic magnetostriction. Magnetostrictive effects in GeCo$_2$O$_4$ are consistent with large magnetostrictive and anisotropic effects that are observed in cobalt compounds because of spin-orbit coupling in high spin octahedral Co$^{2+}$.\cite{slonczewski_1961} The structural distortion of GeCo$_2$O$_4$ at 16\,K gives rise to another discontinuity in cell volume [Fig.~\ref{fig:Structureevol} (b)]. A change in entropy occurs at the magnetic phase transition of GeCo$_2$O$_4$ as illustrated by the nearly coincident anomalies in d$\chi$/d$T$ and the temperature normalized heat capacity [Fig.~\ref{fig:Structureevol} (c)]. Importantly, we note that no additional magnetic or heat capacity anomalies occur at the structural transition temperature of GeCo$_2$O$_4$. This suggests a non-magnetic origin of this distortion. It is likely that entropy changes associated with the structural distortion at 16\,K are concealed in the broad lambda-like heat capacity anomaly of GeCo$_2$O$_4$ that peaks at $\approx$\,22\,K. The temperature normalized heat capacity shows significant entropy changes above $T_N$ due to short range spin correlations in this temperature regime. 

\begin{figure}
\centering \includegraphics[width=3.4in]{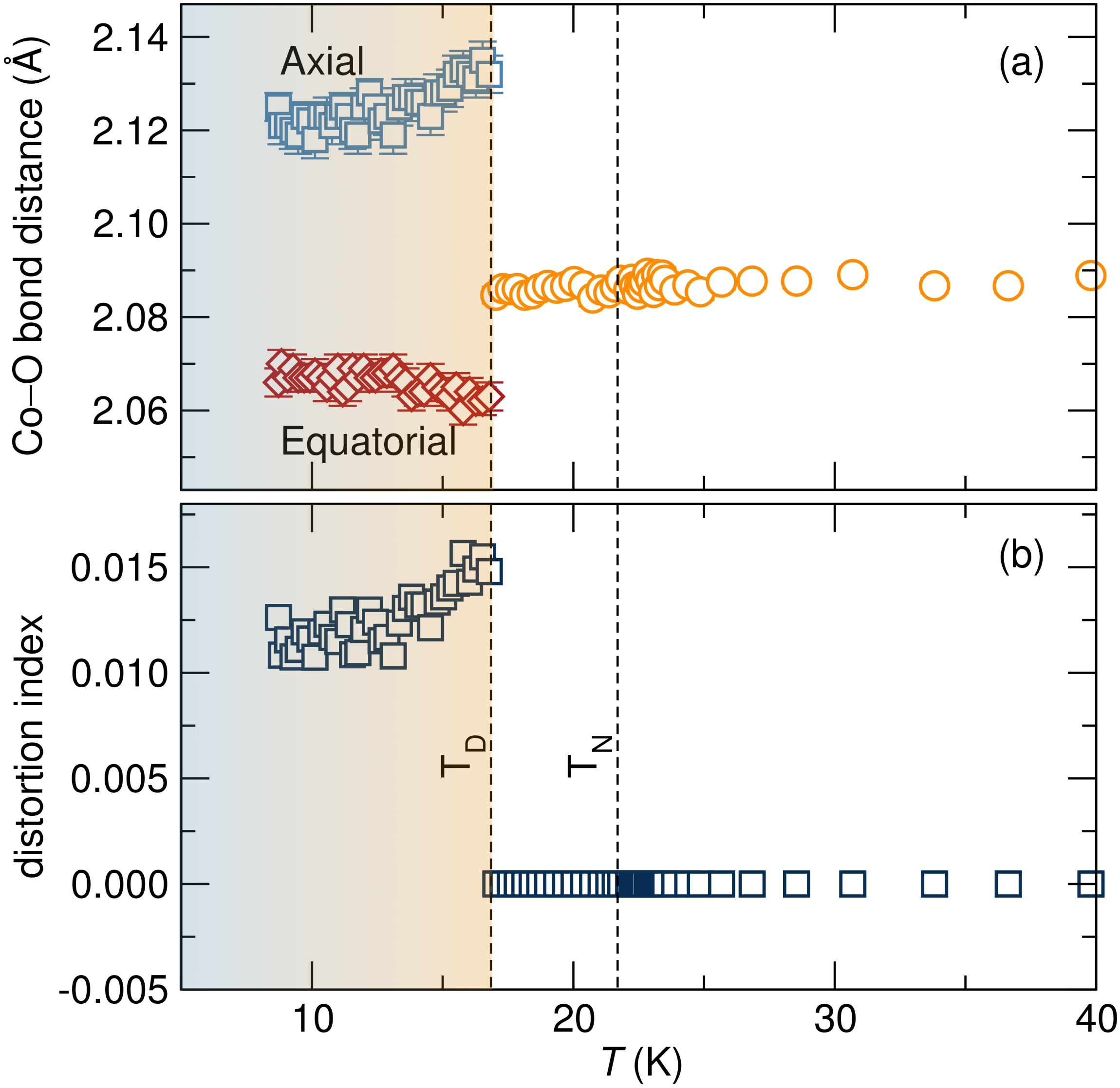}\\
\caption{(Color online) (a) While a single Co--O bond length characterizes CoO$_6$ octahedra in cubic GeCo$_2$O$_4$,  two long Co--O bonds and four short Co--O bonds are observed in the tetragonal phase of GeCo$_2$O$_4$. (b) CoO$_6$ octahedra show no bond distance distortions in the cubic $\textit{Fd}\overline{3}\textit{m}$ phase, however, bond length distortion are observed in the tetragonal $\textit{I}$4$_1$/$\textit{amd}$ phase.
\label{fig:distortions}}
\end{figure}

\begin{figure}
\centering \includegraphics[width=3.4in]{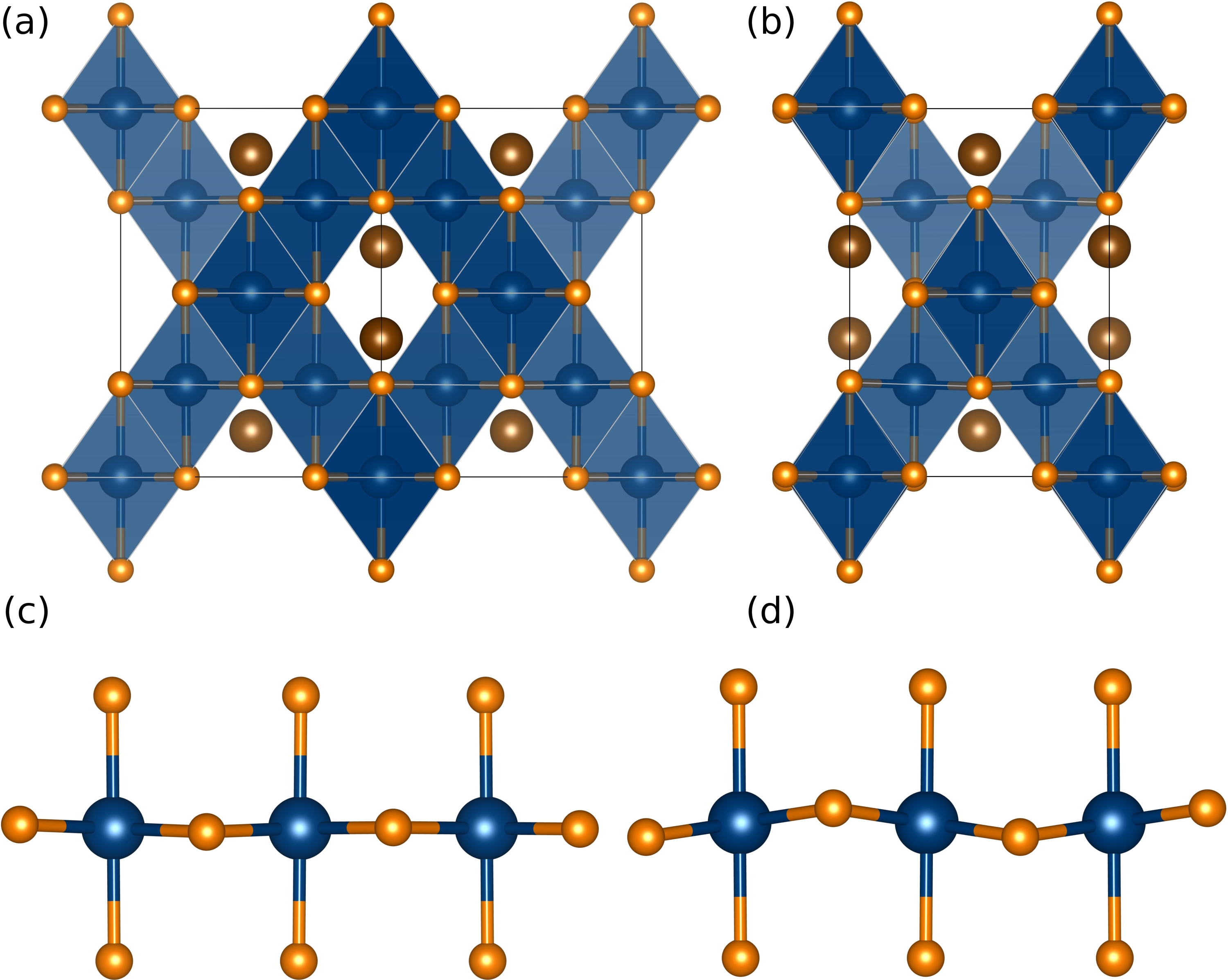}\\
\caption{(Color online) The cubic $\textit{Fd}\overline{3}\textit{m}$ structure of GeCo$_2$O$_4$ at 50\,K and the low temperature tetragonal $\textit{I}$4$_1$/$\textit{amd}$ structure near 8\,K are presented in (a) and (b) respectively. A plane of edge sharing CoO$_6$ octahedra in the cubic $\textit{Fd}\overline{3}\textit{m}$ structure (c) and in the $\textit{I}$4$_1$/$\textit{amd}$ structure near 8\,K (d). The buckling of CoO$_6$ octahedra is enhanced in the tetragonal $\textit{I}$4$_1$/$\textit{amd}$ phase of GeCo$_2$O$_4$, and this likely  occurs to accommodates the elongation of CoO$_6$ octahedra. Distortions in figures (c) and (d) have been enhanced by a factor of 5 to clearly illustrate the structural changes. 
\label{fig:structures}}
\end{figure}

The temperature variation of Ge--O bond distances reveals no bond distance distortions in the cubic or tetragonal phases of GeCo$_2$O$_4$. As a result, in both the cubic and tetragonal phases of GeCo$_2$O$_4$, GeO$_4$ tetrahedra are described by a single bond length. CoO$_6$ octahedra are characterized by a single Co--O bond length in the cubic phase, however, an elongation of CoO$_6$ octahedra is observed in the tetragonal phase [Fig. \ref{fig:distortions} (a)]. While Jahn-Teller effects are expected to be quenched in GeCo$_2$O$_4$ due to strong spin-orbit coupling in Co$^{2+}$,\cite{kanamori_1957} the elongation of CoO$_6$ octahedra is consistent with a weak Jahn-Teller distortion that lifts orbital degeneracy by stabilizing the $xz$ and $yz$ orbitals of the $t_{2g}^5$ states. Bond distance distortions arising from the elongation of CoO$_6$ octahedra in the tetragonal phase are shown in Fig. \ref{fig:distortions} (b). The distortion index $D$ is defined as $D\,=1/n\sum_{i=1}^n(|l_i-\overline{l}|)/(\overline{l})$ where $l_i$ is a given Co--O bond length and $\overline{l}$ is the average Co--O bond length. Figure \ref{fig:structures} shows the cubic and tetragonal structures of GeCo$_2$O$_4$. The elongation of CoO$_6$ octahedra in the tetragonal phase yields an enhanced buckling of Co--O bonds [Fig. \ref{fig:structures} (c) and (d)].

\begin{figure}
\centering
\includegraphics[width=3.4in]{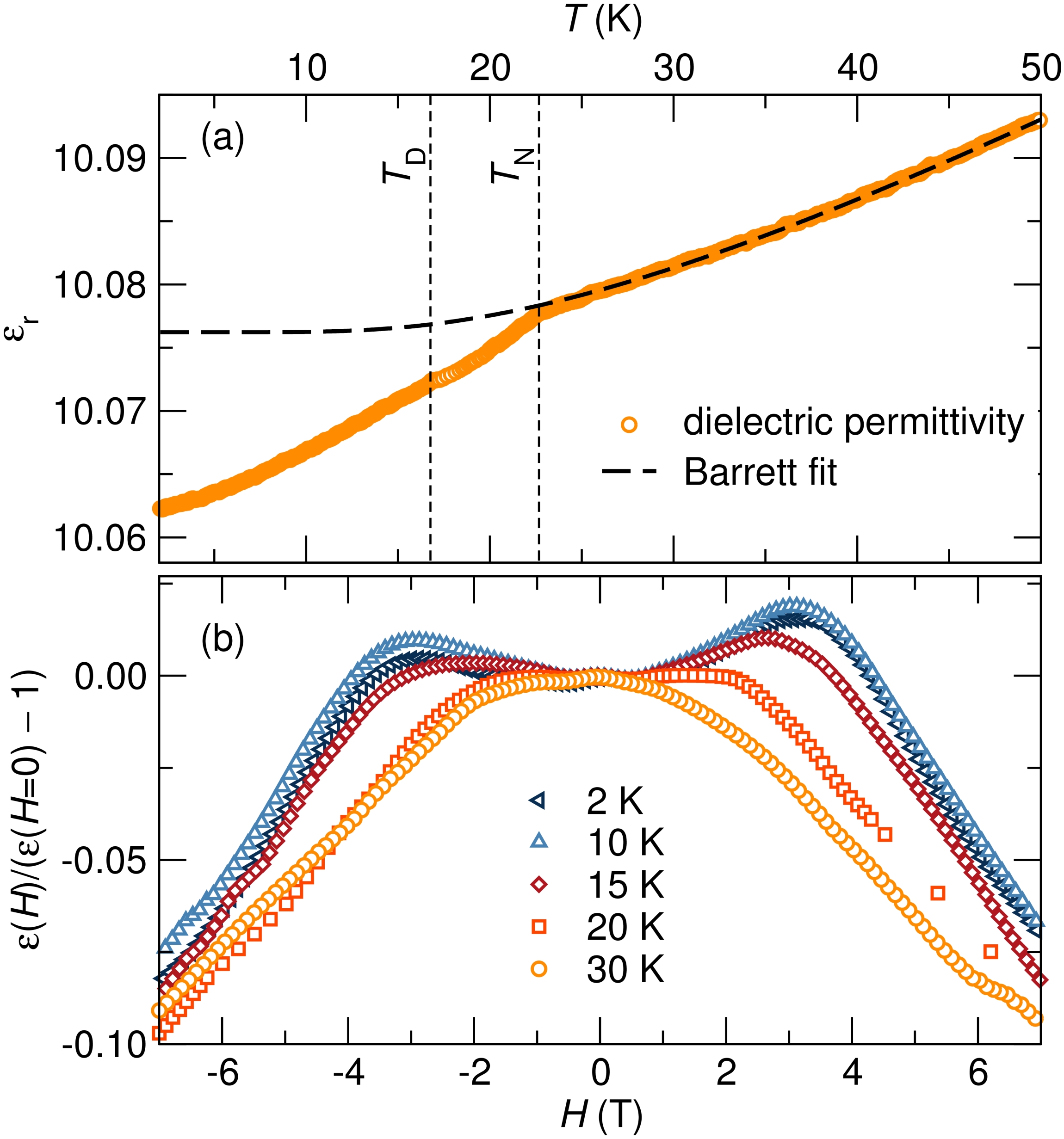}\\
\caption{(Color online) (a) The temperature dependence of the dielectric permittivity of GeCo$_2$O$_4$ shows a dielectric anomaly at the N\'eel temperature ($T_N$\,=\,21\,K). A slight change in slope of the temperature dependent dielectric constant is observed at the structural distortion temperature ($T_D$\,=\,16\,K). The Barrett fit models the dielectric permittivity well above $T_N$, however, the dielectric constant deviates from the Barrett function fit below $T_N$. (b) Relative changes in the dielectric constant of GeCo$_2$O$_4$ measured at 20\,kHz as a function of magnetic field at different temperatures. A distinct change in the field dependence is observed beneath $T_{\rm{N}}$~=~21\,K.}
\label{fig:e-H}
\end{figure}

Concurrent with the magnetic transition of GeCo$_2$O$_4$ is the onset of magnetodielectric behavior. The dielectric permittivity, $\epsilon_r$, is calculated from the capacitance measured in a parallel plate geometry by $\epsilon_r$ = $Cd$/$A$. A suppression of the dielectric constant of GeCo$_2$O$_4$ occurs below $T_N$~=~21\,K as illustrated in Fig.~\ref{fig:e-H} (a), pointing to the magnetic origin of this dielectric anomaly. The structural distortion at $T_D$ leaves a signature in the temperature-dependent dielectric permittivity which shows a change in slope at 16\,K [Fig. \ref{fig:e-H} (a)]. The lattice dielectric constant is modeled by a modified Barrett equation in the temperature range 25\,K $< T <$ 80\,K. The Barrett fit models the dielectric permittivity in the absence of magnetodielectric effects. The dielectric constant at $T$~=~2\,K is 0.057\,\% less than expected by the Barrett function, while the change in sample volume across the transition, as measured by powder synchrotron X-ray diffraction, is only 0.01\,\%. Thus, the change in geometry cannot be fully responsible for the observed deviation in dielectric response. Instead, this difference, whose magnitude is similar to that found in other antiferromagnetic spinels such as Mn$_3$O$_4$,\cite{Tackett_PRB07} is likely due to a magnetodielectric effect. The frequency dependence of the dielectric properties was investigated from 1\,kHz to 20\,kHz, however we did not detect any significant differences in the temperature evolution or magnitude, nor did we see relaxation effects, with tan($\delta$)~$<$~0.0003 for the temperatures and frequencies measured. These observations suggest that the dielectric response is not of magnetoresistive origin, and instead supports the presence of magnetodielectric coupling in GeCo$_2$O$_4$.\cite{Catalan_APL06} 


The dielectric constant can be generally related to optical phonons and their frequencies by the Lyddane-Sachs-Teller relationship. It is possible to more directly connect $\epsilon_r$ to the relevant transverse-optical modes using a Barrett function, as for example, was done for BaMnF$_4$\cite{Fox_PRB80} and MnO,\cite{Seehra_PRB81} and more recently for TbFe$_3$(BO$_3$)$_4$.\cite{Adem_PRB10} The Barrett function is $\epsilon$($T$) = $\epsilon$(0) + $A$/[exp($\hbar\omega_0$/$k_{\rm{B}}T$)$-$1], where $A$ is a coupling constant and $\omega_0$ is the mean frequency of the final states in the lowest-lying optical phonon branch. The refined parameters of the fit are $\epsilon$(0) = 10.0762, $A$ = 0.0626, and $\omega_0$ = 339\,cm$^{-1}$. This $\omega_0$, which is an average, is near the 302\,cm$^{-1}$ value of a transverse-optical phonon $E_g$ mode found by Raman spectroscopy,\cite{Koningstein_JCP72} and suggests a possible spin-phonon coupling mechanism.

Further evidence that this dielectric behavior is magnetic in nature is observed in capacitance measurements performed in a varying magnetic field [Fig.~\ref{fig:e-H}(b)]. We plot the magnetic-field dependent dielectric permittivity with respect to the zero field permittivity using the equation 
$\Delta\epsilon$ =$\epsilon$($H$)/$\epsilon$(0)$-$1. As expected for a magnetodielectric, the field dependent dielectric permittivity changes below $T_{\rm{N}}$. When $T > T_{\rm{N}}$, the observed dielectric response is positive at low applied fields but becomes negative at higher fields. The transition between positive and negative responses occurs at $H$~=~0.5\,T for $T$ = 20\,K and increases to $H$~=~3\,T for $T$ = 2\,K. The magnitude of the positive upturn increases with decreasing temperature until below 10\,K at which point the response begins to weaken. The asymmetry in the field-dependence in positive and negative fields is the result of magnetic hysteresis in the small Co$_{10}$Ge$_3$O$_{16}$ impurity.\cite{Barton_PRB13} The qualitative change in $\epsilon-H$ behavior with temperature suggests that there is substantial magnetodielectric coupling in this system. The changes in the dielectric permittivity in an applied field above $T_{\rm{N}}$ also occur in other antiferromagnetic magnetodielectrics and are not due to magnetodielectric effects.\cite{Tackett_PRB07,Adem_PRB10} Capacitance measurements revealed no magnetodielectric effects in GeNi$_2$O$_4$, however GeFe$_2$O$_4$ was not characterized.



\subsection*{Jahn-Teller degeneracy and spin orbit coupling in GeCo$_2$O$_4$}

The origin of structural transitions in systems with degenerate $t_{2g}$ states that are more than half occupied are difficult to identify. The poor understanding of these distortions arises from the intricate interplay between spin, orbital, and lattice degrees of freedom. In understanding the structural transformation of GeCo$_2$O$_4$, it is illuminating to consider the related binary oxide CoO. CoO has a rocksalt crystal structure and Co$^{2+}$ occupy octahedral sites and have the high spin $3d^7$ electronic configuration of S\,=\,$\frac{3}{2}$ and L\,=\,3 observed in GeCo$_2$O$_4$. CoO exhibits a structural distortion at its N\'eel temperature, $T$\,=\,290\,K. The origin of the structural distortion of CoO is under debate with some reports attributing it to spin-orbit coupling magnetostrictive effects\cite{jauch_2001,goodeneough_1963} while others propose Jahn-Teller ordering.\cite{jauch_2002,ding_2006} A spin-orbit mediated structural distortion can arise from the significant spin-orbit energy $\lambda \textbf{L}\cdot\textbf{S}$ that is equal to or greater than the Jahn-Teller stabilization in high-spin octahedral $3d^7$ systems.\cite{goodeneough_1963} The structural distortion of CoO leads to a compression of CoO$_6$ octahedra; this distortion does not lift spin degeneracy in this material where the $yz$ and $xz$ orbitals remain degenerate.\cite{dunitz_1957,jauch_2002} However, recent high pressure experiments by Ding $et\,al.$ have noted a decoupling of the structural and magnetic ordering in CoO under pressure; magnetic ordering occurring at higher temperatures without an accompanying lattice distortion.\cite{ding_2006} In light of these findings, Ding $et\,al.$ propose a Jahn-Teller mediated structural distortion in CoO which is suppressed under pressure resulting in the onset of antiferromagnetic order without an accompanying structural distortion. The same complexities in identifying the deformation mechanism in CoO are to be expected in GeCo$_2$O$_4$.

There are three main kinds of structural distortions in magnetic spinels. First, there are Jahn-Teller distortions that break orbital degeneracy as observed in CuCr$_2$O$_4$.\cite{dunitz_1957} Jahn-Teller distortions typically occur at temperatures much higher than magnetic ordering temperatures.\cite{dunitz_1957} Then there are magnetostructural transformations where the onset of magnetic order changes the crystal symmetry as reported in NiCr$_2$O$_4$ and CuCr$_2$O$_4$.\cite{Suchomel_PRB12} Finally, there are spin-Jahn-Teller distortions that break the degeneracy in spin configurations, for example in ZnCr$_2$O$_4$ and MgCr$_2$O$_4$.\cite{Kemei_JPCM13} Like magnetostructural distortions, spin-Jahn-Teller transformations occur at the magnetic ordering temperature. Magnetostructural and spin-Jahn-Teller distortions usually involve small distortions of the lattice compared to Jahn-Teller distortions.

Previous studies of the structure and magnetism of GeCo$_2$O$_4$ have associated its structural distortion to magnetostrictive effects\cite{Hoshi_JMMM07} that are present in octahedral Co$^{2+}$ due to degenerate $t_{2g}$ states.\cite{slonczewski_1961} The 1.001 $c/a$ tetragonal elongation measured in GeCo$_2$O$_4$ below 10\,K compares well with spin driven distortions in the geometrically frustrated systems MgCr$_2$O$_4$ and ZnCr$_2$O$_4$.\cite{Kemei_JPCM13}  However, the onset of the distortion below the N\'eel temperature suggests a non-magnetic origin of this lattice distortion. A Jahn-Teller origin of this distortion is plausible given that the deformation can lift spin degeneracy by stabilizing the $xz$ and $yz$ orbitals of the $t_{2g}$ states. Although the small tetragonal distortion of GeCo$_2$O$_4$ is at odds with large Jahn-Teller distortions observed for example in NiCr$_2$O$_4$,\cite{dunitz_1957} a small distortion is expected in degenerate $t_{2g}$ systems due to the weak electronic stabilization achieved through this deformation. The close proximity between the magnetic and structural ordering temperatures is in line with the competition between spin-orbit and Jahn-Teller stabilization. However, spin-orbit coupling is expected to dominate in high spin $3d^7$ complexes\cite{kanamori_1957} and the precise origin of the structural deformation in GeCo$_2$O$_4$ should be further investigated.


The possibility of a gradual deformation that begins at the N\'eel temperature of GeCo$_2$O$_4$ with a broadening of the diffraction reflections and is fully manifested below 16\,K where a splitting of some of the diffraction reflections is observed should also be considered. In this case, the structural transformation would likely be linked to magnetostrictive effects.

\section{Conclusions}

A structural phase transition was observed in the spinel GeCo$_2$O$_4$ at $T_D$\,=\,16\,K using variable-temperature high-resolution synchrotron powder x-ray diffraction and physical property measurements. An analogous transition was not observed in GeFe$_2$O$_4$ or GeNi$_2$O$_4$. Unlike many other magnetic spinels, the magnetic and structural transitions of GeCo$_2$O$_4$ are not coincident and we discuss the decoupling of structural and magnetic ordering in this system considering the effects of magnetostriction and Jahn-Teller ordering. We report the first complete description of the low-temperature $\textit{1}$4$_1$/$\textit{amd}$  crystal structure of GeCo$_2$O$_4$ with $c/a > 1$. In GeFe$_2$O$_4$, we observe a second antiferromagnetic transition, not previously reported, that is reminiscent of GeNi$_2$O$_4$ because it is close in proximity to another transition slightly higher in temperature. Finally, we present evidence for magnetodielectric coupling in GeCo$_2$O$_4$ beneath $T_{\rm{N}}$.\\

\section{Acknowledgments}
We thank Professor Gavin Lawes for preliminary measurements on GeCo$_2$O$_4$ and Dr. John Mitchell for helpful discussions. We thank Dr. Christina Birkel for SPS. This project was supported by the NSF through the DMR 1105301. PTB is supported by the NSF Graduate Research Fellowship Program. MCK is supported by the Schlumberger Foundation Faculty for the Future fellowship. MWG is supported by a NSERC Postgraduate Scholarship and an International Fulbright Science $\&$ Technology Award. We acknowledge the use of MRL Central Facilities which are supported by the MRSEC Program of the NSF under Award No. DMR 1121053; a member of the NSF-funded Materials Research Facilities Network (www.mrfn.org). Use of data from the 11-BM beamline at the Advanced Photon Source was supported by the U.S. Department of Energy, Office of Science, Office of Basic Energy Sciences, under Contract No. DE-AC02-06CH11357. Data were also collected on the ID31 beamline at the European Synchrotron Radiation Facility (ESRF), Grenoble, France. We thank Andy Fitch and Caroline Curfs for providing assistance in using beamline ID31.

\bibliography{GeCo2O4}

\end{document}